# Influence of lattice orientation on growth and structure of graphene on Cu(001)

Joseph M. Wofford,[1,†,*] Shu Nie,[2] Konrad Thürmer,[2] Kevin F. McCarty,[2] Oscar D. Dubon[1,*]

[1] Department of Materials Science & Engineering, University of California at Berkeley, and Lawrence Berkeley National Laboratory, Berkeley, CA 94720, USA

[2] Sandia National Laboratories, Livermore, CA 94550, USA

## Abstract

We have used low-energy electron microscopy and diffraction to examine the significance of lattice orientation in graphene growth on Cu(001). Individual graphene domains undergo anisotropic growth on the Cu surface, and develop into lens shapes with their long axes roughly aligned with the Cu<100> in-plane directions. The long axis of a lens-shaped domain is only rarely oriented along a C<11> direction, suggesting that carbon attachment at "zigzag" graphene island edges is unfavorable. A kink-mediated adatom attachment process is consistent with the behavior observed here and reported in the literature. The details of the ridged moiré pattern formed by the superposition of the graphene lattice on the (001) Cu surface also evolve with the graphene lattice orientation, and are predicted well by a simple geometric model. Managing the kink-mediated growth mode of graphene on Cu(001) will be necessary for the continued improvement of this graphene synthesis technique.

--------------------------------------------------------

## 1. Introduction

Copper foils have proven to be an effective substrate for graphene growth. The (001)-textured Cu foils support the growth of large-area polycrystalline graphene films, which none-the-less remain exclusively one monolayer thick due to the low solubility of C in the metal [1–4]. Recent refinement of this growth process has yielded individual graphene islands exceeding 1 mm in size [5,6]. The scientific and technological significance of these results has made characterizing the fundamental processes behind this synthesis technique an urgent priority.

[†] Current address: Paul-Drude-Institut für Festkörperelektronik, Hausvogteiplatz 5-7, 10117 Berlin, Germany
[*] Corresponding authors:  Tel: +49 30 20377 426     E-mail: joewofford@gmail.com   (Joseph Wofford)
                Tel: 510-643-3851         E-mail: oddubon@berkeley.edu   (Oscar Dubon)

The microstructure of a thin film, or its "quality", is primarily determined by its growth history. Growth, in turn, is driven by the crystal structures of both the film and the substrate. Understanding the relationship between the combined film-substrate crystallography and growth behavior is thus a prerequisite to systematically improve film quality. Here, the crystallography of the graphene/Cu(001) system is defined by the 6-fold symmetric graphene lattice, the 4-fold symmetric Cu facet, and the relative angle between the two. The graphene and Cu lattices are perfectly aligned when a C<10> direction is parallel to either one of two equivalent in-plane orientations, Cu[110] or Cu[1$\bar{1}$0]. However, the lattices of most graphene domains are rotated some degrees away from such ideal alignment, resulting in an entire range of different graphene configurations on the Cu surface [4].

A straightforward examination of the symmetry of the graphene/Cu(001) system begins to reveal how the interplay between crystallography and growth behavior unfolds. The symmetry of the combined structure is determined by which symmetry elements are common to the graphene lattice and Cu surface: a 2-fold rotation. This 2-fold axis manifests itself in the moiré formed by the superposition of the hexagonal graphene lattice on the (001) surface of the Cu. Rather than forming a hexagonal superlattice, as graphene does on a close-packed metal facet [7–11], here the moiré gives rise to a periodic array of parallel ridges in the graphene [12–14]. The combined graphene-Cu(001) symmetry is also reflected in the shape evolution of individual graphene crystals growing on the Cu, where a 2-fold symmetric anisotropic growth rate sculpts many graphene crystals into elongated, lens-like profiles (for an extended discussion please see reference #[4]). The long axes of the lens shaped graphene crystals – which we define as their "fast-growth direction" – align roughly with the Cu<100> in-plane directions, illustrating explicitly the influence of the substrate crystallography.

While this symmetry analysis is useful as a phenomenological guide, it neglects many details of the growth scenario. An exhaustive study that includes the relative orientation of the graphene lattice is necessary to fully understand the significance of crystallography in this growth system. Here, low-energy electron microscopy (LEEM) and diffraction (LEED) were used to study graphene growth by molecular beam epitaxy (MBE). In particular, we examined the influence of the graphene lattice orientation on its fast-growth direction and moiré pattern.

## 2. Experimental methods

Graphene films were synthesized on 25-µm-thick polycrystalline foils (Johnson Matthey, 99.999% Cu, catalog #10950). The foils were annealed at 1000 ºC for 45 minutes in atmospheric pressure Ar-$H_2$ in preparation for C deposition. Growth was performed in the chamber of a LEEM (base pressure ~1 x $10^{-10}$ Torr), where the requisite C flux was generated from an electron-beam-heated graphite rod. The substrate temperature was monitored via a thermocouple welded to the sample mount. Foils were heated to 960 ºC for 10 minutes and subsequently held at 840 ºC for film growth. LEEM was used to continuously monitor the morphology of the sample surface throughout C deposition.

Post-growth analysis using selected-area LEED allowed the crystallography of the graphene domains to be correlated with their growth behavior. The relative orientation between the lattices of the graphene and Cu surface is the smallest angle observed between any set of first-order Cu and graphene diffraction spots (the C<10> to Cu<110> angle). This angle is ≤15º, by symmetry. Furthermore, projecting the fast-growth direction of a graphene domain onto its LEED pattern allows the relationship between growth behavior and crystallography to be directly examined.

## 3. Results

### 3.1 Anisotropic growth

We observed graphene to form polycrystalline, mostly 4-lobed islands on the Cu(001) surface. Each island lobe is a single graphene crystal, and develops with the long axis of its lens shape (i.e., fast-growth direction) roughly along a Cu<100> in-plane direction (see **Figure1.a**). If uninterrupted by surface irregularities, each graphene domain developed this morphology regardless of its lattice orientation.

Previous investigations of graphene growth on Cu(001) have shown how such four-lobed, polycrystalline islands may form. Each of the four constituent domains of the polycrystalline islands shares the same nucleation event despite their different lattice orientations. These large domains developed in a configuration favoring rapid growth, allowing them to expand into elongated lobes at the expense of any less-favorably oriented domains [4]. In the current paper we examine the relationship between the fast-growth direction of graphene domains and their lattice orientation more closely and find that certain combinations of fast-growth direction and lattice orientation occur much less frequently than others.

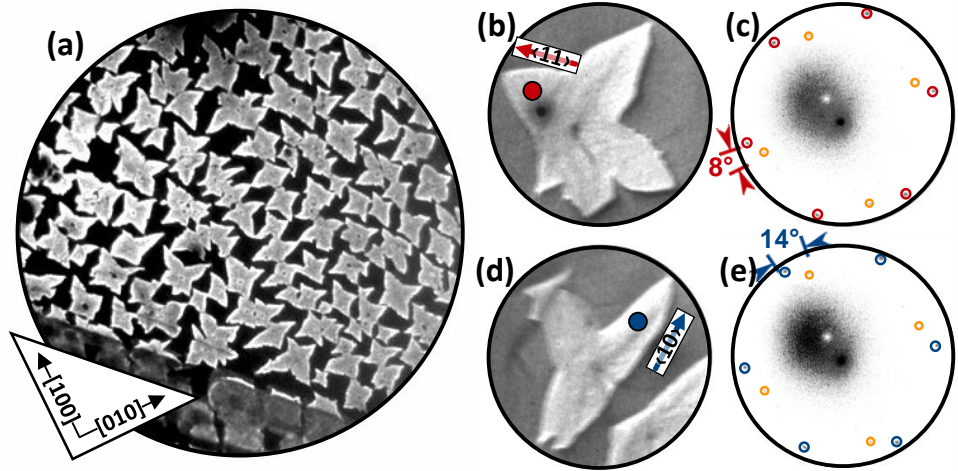

**Figure 1.** LEEM micrographs (a, b, d) and LEED patterns (c, e) examining graphene domains on Cu(001). The red and blue dots in (b) and (d) are the locations where (c) and (e) were collected, respectively. The orange circles in (c) and (e) are centered on Cu diffraction spots, while the red and blue highlight diffraction from graphene. Many graphene islands growing on Cu(001) develop a distinct 4-lobed morphology, where the fast-growth direction of the lobes lies close to a Cu<100> in-plane direction, as is shown by the consistent island-to-island orientation in (a) (FOV = 46 µm). Also in (a) are graphene islands of other shapes, which typically form due to interactions with surface inhomogeneities and other graphene islands. The dark lower-left portion of (a) is a twinned region of the Cu(001) substrate. The lattice of the graphene domain in (b, c) is rotated 8° relative to the Cu surface (see spacer-arrows) in such a way as its fast-growth direction (red arrow in (b)) is closer to a C<11> direction. The graphene domain in (d, e) is rotated by 14° relative to the Cu (spacer-arrows), but has its fast-growth direction (blue arrow in (d)) close to a C<10> direction. Small and moderate lattice rotations, such as that in (b, c) are equally likely to be toward the C<10> and C<11> directions. Larger rotations, such as (d, e), almost always move the fast-growth direction towards a C<10> lattice direction. FOV in (b, d) is 7 µm.

For a graphene domain approaching perfect alignment, i.e. C<01>∥Cu<110>, the fast-growth direction is halfway between the two high-symmetry directions of the graphene lattice, C<10>

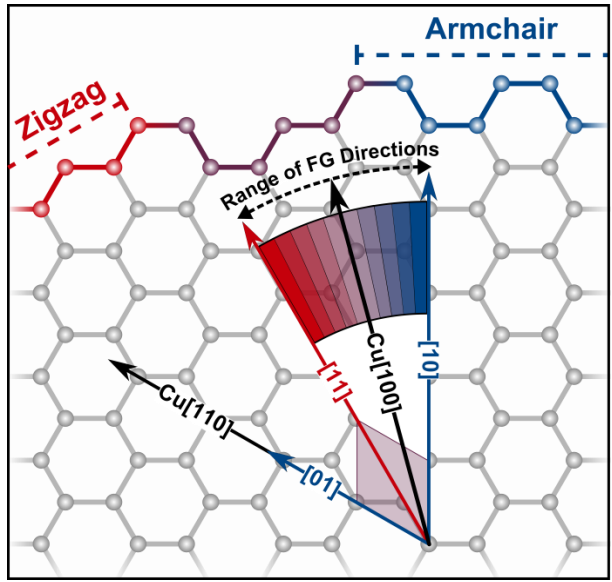

**Figure 2.** Schematic diagram illustrating lattice directions and edges of the graphene lattice (the purple diamond is the graphene unit cell). Because the fast-growth direction remains near a Cu<100> direction, a rotation of the graphene lattice changes the lattice direction along which fast-growth occurs. When the lattice of a graphene domain approaches perfect alignment with the Cu(001) surface (i.e. C[10] ∥ Cu[110] or Cu[1-10], see arrows on left of unit cell) the fast-growth direction is directly between a C<10> and a C<11> direction. Any rotation away from alignment moves the fast-growth direction towards either the C<10> or C<11> direction. A graphene edge which is perpendicular to a C<10> direction (blue) has the armchair structure, while an edge perpendicular to a C<11> direction (red) has the zigzag structure. The purple edge section is a mix of these two high-symmetry edge structures.

and C<11> (see **Figure 2**). The graphene edge perpendicular to the fast-growth direction – that is, the "fast-growth edge" where new carbon is being incorporated most rapidly – is similarly an equal blend of the armchair and zigzag edge configurations. The lattice rotation of a misaligned graphene domain moves the fast-growth direction closer to either a C<10> or C<11> direction, with a corresponding shift in the edge structure (see **Figure 1.b-e** for examples of domains from each category). If fast-growth is closer to a C<11> direction the fast-growth edge has a larger zigzag component, and domains growing fast along C<10> have a larger armchair component in their fast-growth edges. How the graphene lattice is oriented relative to the Cu<100> in-plane direction thus dictates the structure of the graphene fast-growth edge.

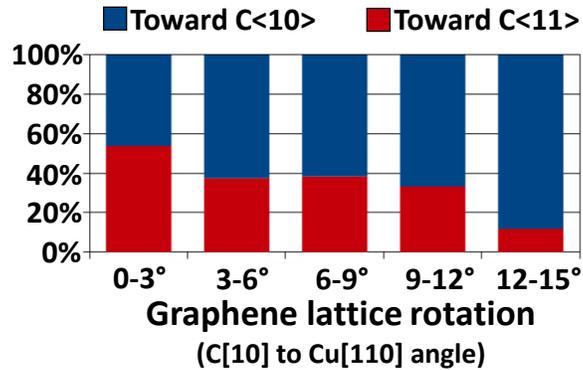

**Figure 3.** The proportion of graphene domains with their fast-growth direction toward a C<10> or C<11> lattice direction, as a function of the graphene rotation angle. For small rotations, fast growth is equally likely to be toward a C<11> or C<10> direction. However, the majority of graphene domains which are misaligned relative to the Cu(001) surface by more than 9º undergo fast-growth near a C<10> lattice direction.

A summary of LEED data, collected from over 130 different individual graphene domains, shows the likelihood that the graphene lattice is rotated such that fast-growth occurs closer to a C<10> or C<11> direction (**Figure 3**). Of those graphene domains that are aligned within ±3º of the Cu surface, approximately an equal number are rotated in either direction. When the lattice misalignment increases to ±3-6º or ±6-9º, ~60% of domains are oriented such that fast-growth occurs closer to a C<10> direction. The distribution increasingly favors the C<10> directions for larger lattice rotations. For instance, approximately 70% of domains misaligned by ±9-12º are rotated such that a C<10> direction is closer to the fast-growth direction, and this proportion jumps to ~90% for domains rotated by ±12-15º. The relative absence of graphene domains with fast-growth occurring along a C<11> direction implies that carbon incorporation becomes increasingly unfavorable as the graphene edge approaches the purely zigzag configuration.

It is interesting to note that graphene lobes which are misaligned such that fast-growth occurs closer to a C<11> lattice direction are often part of a polycrystalline graphene island with an atypical domain structure. In these unusual islands, two neighboring lobes have precisely the same lattice orientation, suggesting they are part of a single large graphene domain which includes both lobes (see **Figure 4**). Because they are both part of the same large domain - and thus have the same lattice orientation - one of the lobes undergoes fast-growth closer to a C<10>

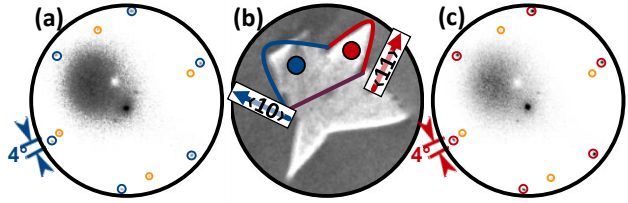

**Figure 4.** LEED patterns (a, c) and a LEEM micrograph (b) examining a graphene island on Cu(001). The blue and red dots in (b) are where (a) and (c) were collected, respectively. The orange circles in (a) and (c) are centered on Cu diffraction spots, while the red and blue highlight diffraction from graphene. Occasionally graphene islands have an atypical domain structure, where two neighboring island lobes are part of a single large domain (see approximate domain outline in (b)). In these circumstances, one lobe is C<10>, armchair-oriented (blue outlined section, blue arrow) while its neighbor is a C<11>, zigzag-tipped lobe (red outlined section, red arrow). The lattice rotation of the graphene domain is 4º, and the FOV in (b) is 7 μm.

direction, while the other is C<11>-oriented. It is possible that the accelerated growth of the C<10>-oriented section of the larger domain stabilized that graphene crystal during the early stages of growth, allowing the unfavorably oriented C<11> section to persist rather than being subsumed by a more preferentially oriented domain. This suggests the inhibition against accelerated growth along the C<11> directions may be even more pronounced than the distribution of graphene orientations reported in **Figure 3** indicates.

3.2 The graphene-Cu(001) moiré

Rotating the graphene lattice also changes the moiré it forms with the Cu surface. Selected-area LEED patterns collected from graphene domains within ±9º of alignment with the Cu(001) substrate (i.e. C<01>∥Cu<110>) often have two diffraction spots in addition to those directly attributable to the C and Cu lattices (**Figure 5**). These extra spots are not observed when the lattice misalignment is larger than ~±9º. The orientation and separation of these diffraction spots vary, and when calibrated with the graphene and Cu patterns suggest a physical phenomenon with a periodicity ranging from 8.4 Å to 12.3 Å. Scanning-tunneling microscopy studies of graphene on Cu(001) have reported the ridged moiré to have periodicities of 11 Å [12], 12±1 Å [13], and 13.5 Å [14], which generally agree with the range seen here. Thus, we attribute the additional diffraction spots to the graphene-Cu(001) moiré.

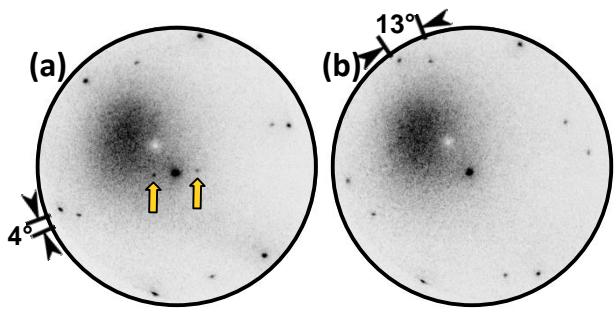

**Figure 5.** LEED patterns from graphene domains grown on Cu(001). Diffraction spots from the ridged graphene-Cu(001) moiré can be seen in (a) (yellow arrows, lattice rotation is 4º). Diffraction from domains rotated by more than ~9º, such as (b), does not show evidence of the ridged moiré (lattice rotation is 13º). The orientation and periodicity of the moiré ridges change with the graphene lattice orientation.

To ascertain how the graphene-Cu(001) moiré evolves as the orientation between the two materials shifts, LEED from graphene domains with many different orientations was analyzed, as is summarized in **Figure 6**. The LEED patterns examined include the full range of graphene orientations with observable moiré diffraction spots, covering a ~18º arc centered on the perfectly aligned configuration. As the graphene lattice swivels from +9º to -9º the orientation of the ridged moiré rotates by ~90º, starting nearly parallel to Cu[010], moving through Cu[110] when the graphene lattice is perfectly aligned, and ending with the ridges roughly parallel to Cu[100]. The periodicity of the ridges varies simultaneously, with the shortest wavelength of 8.4 Å when the moiré is parallel to a Cu<100> direction, and the longest of 12.3 Å when it is parallel to a Cu<110> direction. For a perfectly aligned graphene domain, the ridges are along a zigzag direction; for misalignments of ~7º the ridges are rotated ~37º with respect to Cu[110] and thus nearly parallel to an armchair direction.

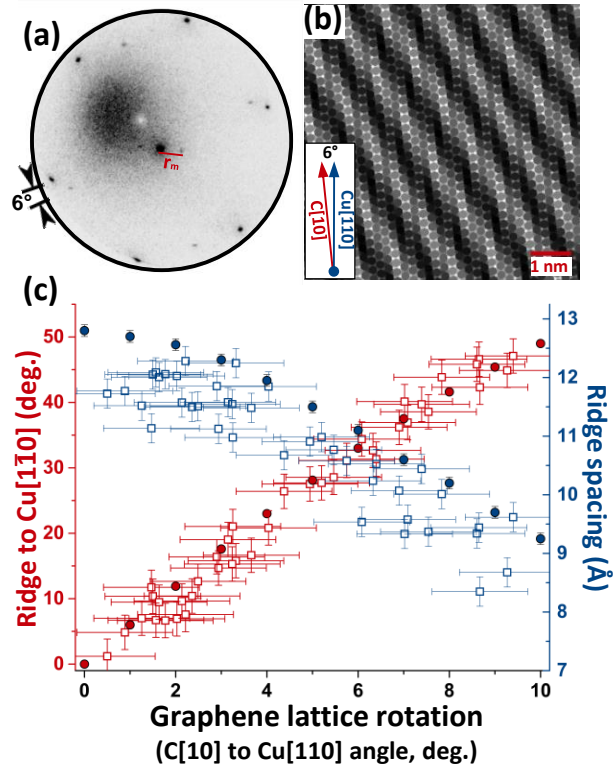

**Figure 6.** LEED from a graphene domain grown on Cu(001) (a) showing the diffraction spots of the ridged graphene-Cu(001) moiré (graphene lattice rotation is 6º). (b) A rendering from a simple geometric model of the graphene-Cu(001) surface, with the graphene lattice rotated by 6º (please see reference #[15] for model details). Brighter atoms are higher. The ridged moiré is clearly visible. (c) A plot summarizing how the graphene-Cu(001) moiré evolves as the graphene lattice rotates, and comparing its orientation and periodicity with those predicted by the model (open squares are from LEED, closed circles from the model). The model accurately recreates the orientation of the ridges, but consistently predicts a larger ridge spacing than is observed by LEED. This may be the result of a dilation in the LEED patterns, which would maintain angular fidelity but lead to consistent underestimation of the ridge spacing.

To determine if the observed orientations and periodicities of the moiré are purely the result of geometry, we compare them with a simple geometric simulation based on laterally unstrained graphene and Cu(001) lattices. As **Figure 6.c** shows, the orientation of the moiré ridges is accurately reproduced (please see reference #[15] for simulation details). However, the spacing between ridges observed by LEED is slightly smaller than predicted by the model. The discrepancy is likely the result of a small dilation in the LEED patterns near the specular beam. This would introduce systemic error to the measured ridge spacing while preserving angular fidelity, thus not altering the observed ridge orientation. The good agreement between observation and model suggest that the details of the ridged graphene-Cu(001) moiré are a straightforward result of the geometry of the two lattices.

## 4. Discussion

We have observed that although graphene on Cu(001) undergoes attachment-limited, anisotropic growth regardless its lattice orientation, certain orientations of graphene occur less often than others. It is much more common for the fast-growth direction of a highly misaligned graphene domain to be close to a C<10> lattice direction than a C<11> direction, which suggests carbon adatom incorporation at zigzag graphene edges is unfavorable. By further examining the observed distribution of graphene orientations in the context of attachment-limited growth, we are able to gain additional insight into the relationship between graphene lattice orientation and growth behavior. As on many metal surfaces, graphene growth on Cu(001) is an attachment limited process, meaning that the incorporation of carbon adatoms at the edge of the graphene sheet is the rate limiting step in the expansion of the graphene crystal [4,16–18]. Thus, the specific atomic geometry at the growth front provides insight into the atomic nature of the growth. For instance, the attachment-limited growth of graphene on Ru(0001) is mediated by the incorporation of multi-atom carbon clusters at majority zigzag edges. Each cluster attachment event creates two kinks in the zigzag edge, which then act as attachment sites for further carbon adatom incorporation (see **Figure 7**) [16]. Similarly, CVD experiments on Pt foil substrates have shown a strong correlation between the line-density of kinks in zigzag graphene edges and their growth and etching rates [19]. The increased likelihood of accelerated growth occurring along the C<10> directions is consistent with the armchair edge structure serving a similar function in graphene growth on Cu(001). Conversely, as the graphene edge approaches a purely zigzag structure there are fewer available adatom attachment sites and growth is inhibited. The kink-mediated adatom attachment process suggested by this interpretation is consistent with recent simulations of graphene growth on Cu(001), and other metals [20,21].

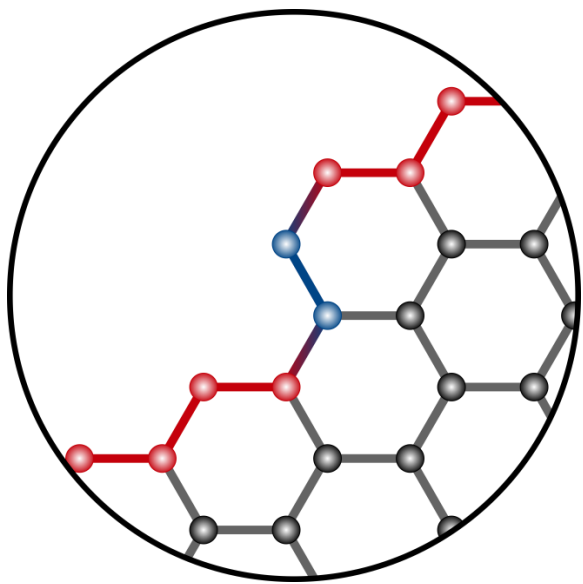

**Figure 7.** A schematic diagram of an armchair-step, or kink (blue), in a majority-zigzag graphene edge (red). Such kinks may facilitate adatom attachment during graphene growth on Cu(001).

We may also look to ambient pressure CVD and seeded graphene growth on Cu(001) for insight into the role of kinks in mediating adatom attachment. This growth process can yield hexagonal, single-domain islands which are bound by zigzag edges [22]. Because kinetically determined crystal shapes are typically terminated by their slowest growing facets [23], this supports the conclusion that carbon attachment is inhibited at purely zigzag graphene edges. We emphasize that regardless of the graphene lattice orientation it is the Cu surface which dictates the possible fast-growth directions. So, while kink-mediated adatom attachment is consistent with the observed growth behavior, it can offer only a partial explanation.

The ridged moiré formed by graphene may also factor into its attachment-limited, anisotropic growth mode on Cu(001), although this is unlikely. A rotation of the graphene lattice such that C[10] is closer to the fast-growth direction causes the moiré ridges to simultaneously rotate away from the fast-growth direction, eventually becoming perpendicular once the graphene lattice is rotated by ~9º. The opposite is true for rotations towards C[11], with the ridges eventually becoming parallel to the fast-growth direction. If the moiré ridges persist to the perimeter of the graphene sheet they would further differentiate the atomic geometry at the growth front. However, the moiré pattern was only detected in graphene domains within ±9º of alignment with the Cu surface, where an approximately equal number of domains are rotated towards the C<10> and C<11> directions. This evidence supports the conclusion that it is the change in edge structure caused by a lattice rotation, and not the associated change in the moiré pattern, which reduces the likelihood for accelerated growth along the C<11> directions.

## 5. Summary

In summary, we have used LEEM and LEED to examine how the growth and structure of graphene on Cu(001) are affected by the relative orientation of the two lattices. The superposition of the hexagonal graphene lattice on the (001) Cu surface results in a ridged moiré, where the orientation and spacing of the ridges depend on the relative graphene orientation. The orientation of the graphene lattice also determines the structure of the graphene edge perpendicular to the Cu<100>, fast-growth direction. Rarely do we find graphene domains with fast-growth occurring close to a C<11> direction, suggesting that adatom incorporation is mediated by kinks, and is thus less likely at purely zigzag graphene edges. Although this interpretation is consistent with the relevant literature, it does not elucidate the mechanism through which fast-growth is restricted to the Cu<100> in-plane directions. Identifying the importance of kinks in the attachment limited growth of graphene on Cu(001) is none-the-less an important step in harnessing the technological potential of this growth method.

**Acknowledgements**

This work was supported by the NSF under Grant No. DMR-1105541 (ODD and JMW) and by the Director, Office of Science, Office of Basic Energy Sciences, Division of Materials Sciences and Engineering, of the U.S. Department of Energy Contract No. De-Ac04-94AL85000 (SN, KT, and KFM).